\documentclass[11pt,a4paper]{article}

\usepackage[margin=2.6cm]{geometry}
\usepackage{amsmath,amssymb,amsthm,mathtools}
\usepackage{braket}
\usepackage{microtype}
\usepackage[hidelinks]{hyperref}
\usepackage{ulem}[normalem]
\usepackage[nameinlink,capitalize]{cleveref}
\usepackage{authblk}
\usepackage{xcolor}

\title{An Operational Resolution of the Third-Particle Paradox\thanks{The first version of these notes was written in April 2020 and has been circulated since then. The present paper gives them in a slightly adapted form. The findings were also presented in the talk ``\textit{The Common Misunderstanding about Quantum Reference Frames and Causal Orderings},'' given by {\v C}. Brukner at RQI North 2024, the 14th Annual Conference on Relativistic Quantum Information, held on 6 August 2024 in Prague, Czech Republic.}}
\author[1,2]{{\v C}aslav Brukner}
\author[1]{Esteban Castro-Ruiz}
\author[4]{Marius Krumm}
\affil[1]{Institute for Quantum Optics and Quantum Information - Vienna, Austrian Academy of Sciences, Vienna, Austria}
\affil[2]{Faculty of Physics, University of Vienna, Vienna, Austria}
\affil[4]{University of Innsbruck, Department of Theoretical Physics, Technikerstr. 21a, A-6020 Innsbruck, Austria}
\date{}

\newcommand{\cH}{\mathcal H}

\newcommand{\cM}{\mathcal M}
\newcommand{\cA}{\mathcal A}
\newcommand{\tr}{\operatorname{Tr}}
\newcommand{\Id}{\mathbb{I}}
\newcommand{\End}{\operatorname{End}}

\begin{document}
\maketitle

\begin{abstract}
We give an operational resolution of the third-particle paradox, relevant in the theory of quantum reference frames. The apparent paradox is that a system which is irrelevant in one quantum-reference-frame description can seem to become relevant after changing to another quantum reference frame, because the reduced state obtained after transforming a larger system need not agree with the state obtained by first discarding the extra system and then transforming. We argue that this comparison is not operationally meaningful unless the observables are transformed together with the states, or equivalently, unless the subsystem that needs to be discarded is properly identified. If the third particle is irrelevant for all measurements actually available in the original frame, then the transformed measurements form a restricted algebra in the new frame for which the third particle remains irrelevant. The paradox therefore results from replacing an operational statement about probabilities by a stronger, representation-dependent statement about equality of reduced density operators. We close by relating the question of when degrees of freedom may be discarded to the observable-induced, operational approach to subsystem structure.
\end{abstract}

\section{Introduction}

Quantum reference frames (QRFs) extend the ordinary notion of a reference frame by treating the system that serves as a frame as a quantum system, which may itself be in a superposition or entangled with other systems~\cite{AharonovKaufherr1984,Rovelli1991,BartlettRudolphSpekkens2007,Giacomini2019,Vanrietvelde2020}. One of the main features of QRF transformations is that they may change the factorisation of the Hilbert space of the systems external to the reference frame, thereby changing their coherence and entanglement properties~\cite{Giacomini2019,Vanrietvelde2020,deLaHametteGalley2020, ali2022quantum, CastroRuizOreshkov2025}. This idea is central to modern approaches to QRF transformations and to the search for a relational formulation of quantum physics.

The third-particle paradox was first stated by Angelo, Brunner, Popescu, Short and Skrzypczyk in their analysis of physics within a QRF~\cite{Angelo2011}. In their formulation, two particles may display coherence relative to one quantum frame, while the mere inclusion of a third, apparently irrelevant particle can change the relative description. A related version of the puzzle appears in the discussion of non-inertial quantum frames by Angelo and Ribeiro~\cite{AngeloRibeiro2012}. The problem has subsequently been revisited from the viewpoint of different formalisms for QRF transformations. In formulations closely related to the perspective-neutral framework~\cite{Vanrietvelde2020}, Krumm, H{\"o}hn and M{\"u}ller argue that the puzzle is tied to the non-uniqueness of embedding fewer relational systems into more relational systems, and introduce a relational trace as a formal resolution~\cite{Krumm2021}. This analysis was extended in the finite Abelian group setting by H{\"o}hn, Krumm and M{\"u}ller~\cite{HoehnKrummMueller2022}. More recently, inspired by compositional considerations, Castro-Ruiz and Oreshkov developed a relative-subsystems approach in which relative subsystems are defined independently of any additional subsystem one may consider, thus formulating an alternative resolution of the paradox~\cite{CastroRuizOreshkov2025}. The present note gives a complementary operational resolution: it asks not only which states, but also \textit{which measurements}, are actually being compared. We base our analysis in the so-called ``perspectival'' approach to QRF transformations, which defines QRF transformations based only on the operational perspective of observers ``attached'' to such QRFs, without introducing any external or global structure. Nevertheless, our main conclusion -- that solving the paradox requires an operational identification of the subsystem structure -- is compatible with the resolutions of~\cite{Angelo2011, Krumm2021, CastroRuizOreshkov2025}. In this respect, our resolution is close in spirit to the operational view of subsystem structure, according to which the accessible observables help determine the relevant factorisation~\cite{Zanardi2001,ZanardiLidarLloyd2004}.

The key point is simple. A physical claim that a distant third system can be ignored should not be interpreted naively as a statement about the formal equality of density operators with and without the ``third system''. Rather, it is a statement about \textit{the probabilities of a specified class of measurements}. If the original observer has access only to measurements on systems $A$ and $B$, relative to their own QRF $C,$ and the third system $D$ is irrelevant for those measurements, then the operational content is
\begin{equation}\label{eq:operational-equivalence-C}
 \tr\!\left[\rho^{(C)}_{ABD}\left(O^{(C)}_{AB}\otimes \mathbf{1}_D\right)\right]
 =
 \tr\!\left[\rho^{(C)}_{AB}O^{(C)}_{AB}\right]
\end{equation}
for all observables $O^{(C)}_{AB}$ in the accessible set. A change of QRF must transform not only states but also the observables whose statistics define the operational situation. Once this is done, the third system remains irrelevant for precisely the transformed operational question.

\section{The third-particle paradox}

Consider a description from the quantum reference frame $C$. Systems $A$ and $B$ are the systems of interest. A further system $D$ is far away, or otherwise irrelevant, so that one may either include it in the state description or not. We write the corresponding states as
\begin{equation}
  \sigma^{(C)}_{ABD}\in\End(\cH_A\otimes\cH_B\otimes\cH_D),
  \qquad
  \rho^{(C)}_{AB}\in\End(\cH_A\otimes\cH_B,),
\end{equation}
where $\mathrm{Tr_D \,\sigma^{(C)}_{ABD} = \rho^{(C)}_{AB}.}$ According to Ref.~\cite{Giacomini2019}, a change of perspective from $C$ to $A$ is represented by a unitary QRF transformation, denoted schematically by $S^{(\text {no D})}_{C\to A}$ in the description without $D$, and $S^{(\text{with D})}_{C\to A}$ in the description with $D$. Applying it to the description in which $D$ is included gives
\begin{equation}\label{eq:state-transform-large}
 \sigma^{(A)}_{CBD}
 =
 S^{(\text {with D})}_{C\to A}\, \sigma^{(C)}_{ABD}\,S^{(\text {with D})\dagger}_{C\to A}.
\end{equation}
If $D$ is not included, one obtains instead
\begin{equation}\label{eq:state-transform-small}
 \rho^{(A)}_{CB}
 =
 S^{(\text {no D})}_{C\to A}\,\rho^{(C)}_{AB}\, S^{(\text {no D})\dagger}_{C\to A}.
\end{equation}

The apparent paradox is the observation that, in general,
\begin{equation}\label{eq:paradox}
 \rho^{(A)}_{CB}\neq \tr_D\sigma^{(A)}_{CBD}.
\end{equation}
This paradox is caused by the issue that $S^{(\text {with D})}_{C\to A}$ does not factor out $D$, i.e.
\begin{equation}
    S^{(\text {with D})}_{C\to A} \ne S^{(\text {no D})}_{C\to A} \otimes S_D.
\end{equation} 
At first sight this seems to mean that a mere decision to include or exclude an irrelevant third particle in the original description changes the physics seen from another QRF. If this were the correct conclusion, then one might worry that a consistent and meaningful description of subsystems relative to quantum reference frames in not viable. 

We now argue that this conclusion is not correct. The original operational claim did not say that $D$ is irrelevant for every conceivable observable relative to $A$. It said only that $D$ is irrelevant for the actual measurements considered relative to the original frame $C$. Hence, relative to $C$, these measurements were restricted to systems $A$ and $B$ only; otherwise, the inclusion or exclusion of $D$ could not be operationally irrelevant in the first place. 

\section{Operational equivalence in the original frame}

Let $\cM_C$ be the set of effects that are operationally available in the description from $C$ and that concern only the systems $A$ and $B$. The assertion that $D$ may be ignored from the perspective of $C$ means that, for every $O^{(C)}_{AB}\in\cM_C$,
\begin{equation}\label{eq:equiv-C}
 \tr\!\left[\sigma^{(C)}_{ABD}\left(O^{(C)}_{AB}\otimes \mathbf{1}_D\right)\right]
 =
 \tr\!\left[\rho^{(C)}_{AB}O^{(C)}_{AB}\right].
\end{equation}
Eq.~\eqref{eq:equiv-C} is a statement about a restricted operational algebra. The third system is irrelevant relative to the pair
\begin{equation}
  \left(\sigma^{(C)}_{ABD},\,\{O^{(C)}_{AB}\otimes\mathbf{1}_D:O^{(C)}_{AB}\in\cM_C\}\right).
\end{equation}
Thus, the phrase ``$D$ need not be included'' is shorthand for equality of all probabilities generated by the accessible measurements, not for an unrestricted equality of descriptions.

\section{Changing perspective: transform states \textit{and} observables}

A passive change of QRF maps both states and effects. Therefore, when $D$ is included in the description, the effects corresponding to
$O^{(C)}_{AB}\otimes\mathbf{1}_D$ in the perspective of $A$ are
\begin{equation}\label{eq:transformed-observable-large}
 \widetilde O^{(A)}_{CBD}
 =
 S^{(\text{with D})}_{C\to A}\left(O^{(C)}_{AB}\otimes\mathbf{1}_D\right)S^{(\text{with D})\dagger}_{C\to A},
\end{equation}
while the corresponding effects in the description where $D$ was never included are
\begin{equation}\label{eq:transformed-observable-small}
 O^{(A)}_{CB}
 =
S^{(\text{no D})}_{C\to A}O^{(C)}_{AB}S^{(\text{no D})\dagger}_{C\to A}.
\end{equation}

Because the transformation is unitary, and hence preserves probabilities, Eq.~\eqref{eq:equiv-C} immediately implies
\begin{equation}\label{eq:equiv-A}
 \tr\!\left[\sigma^{(A)}_{CBD}\widetilde O^{(A)}_{CBD}\right]
 =
 \tr\!\left[\rho^{(A)}_{CB}O^{(A)}_{CB}\right]
\end{equation}
for all $O^{(C)}_{AB}\in\cM_C$.

This is the operational resolution. The transformed set
\begin{equation}\label{eq:transformed-algebra}
 \widetilde{\cM}_A
 :=
 \left\{
 S^{(\text{with D})}_{C\to A}\left(O^{(C)}_{AB}\otimes\Id_D\right)S^{(\text{with D})\dagger}_{C\to A}
 :O^{(C)}_{AB}\in\cM_C
 \right\}
\end{equation}
is generally not the full algebra of observables on $C$, $B$, and $D$ from the $A$-perspective. It is only the image of the measurements that were actually under consideration in the $C$-perspective. For this restricted set, the third system remains irrelevant. There is no operational difference between including or excluding $D$ as long as the same physical measurements are compared or, in other words, as long as the operationally meaningful subsystem is identified.

However, nothing prevents an observer associated with the $A$-perspective from performing more general measurements than those in $\widetilde{\cM}_A$. Such measurements may genuinely require a description including $D$. But that is a different operational scenario. It corresponds to enlarging the measurement set after changing perspective, not to passively redescribing the original experiment.

\section{Discarding superficial degrees of freedom}

The preceding discussion leaves a natural question. Given the pair
\[
 \left(\rho^{(A)}_{CBD},\widetilde{\cM}_A\right),
\]
how can one recognise that some degrees of freedom are superficial, and how
can one discard them? The point is that superficiality is not defined by the
subsystem label $D$ alone, but by the accessible algebra of observables. Let
\[
 \cA_A:=\operatorname{alg}(\widetilde{\cM}_A)
\]
be the algebra generated by the transformed observables. A degree of freedom
is superficial relative to $\cA_A$ if all observables in $\cA_A$ act trivially
on it, so that no measurement in the accessible set can distinguish changes in
that degree of freedom.

Thus the appropriate reduction is not necessarily the ordinary partial trace
over $D$. Rather, one seeks a reduced description preserving all accessible
expectation values:
\begin{equation}\label{eq:operational-reduction}
 \tr\!\left[\rho^{(A)}_{CBD} X\right]
 =
 \tr\!\left[\rho_{\mathrm{red}}\,\pi_{\mathrm{red}}(X)\right],
 \qquad
 X\in\cA_A .
\end{equation}
This is the operational meaning of tracing out superficial degrees of freedom.

In the present case the accessible algebra has the special form
\[
 \cA_A
 =
 S^{(\text{with D})}_{C\to A}
 \left(
   \cA_C\otimes \Id_D
 \right)
 S^{(\text{with D})\dagger}_{C\to A},
\]
where $\cA_C$ is generated by the original observables on $A$ and $B$. Hence,
before the QRF transformation, the algebra acts trivially on $D$. The
algebra-adapted way of discarding the superficial degree of freedom is
therefore to undo the QRF transformation, trace out $D$ in the original
description, and then transform the reduced description:
\[
 \rho^{(A)}_{CBD}
 \longmapsto
 S^{(\text{no D})}_{C\to A}\,
 \tr_D\!\left[
   S^{(\text{with D})\dagger}_{C\to A}
   \rho^{(A)}_{CBD}
   S^{(\text{with D})}_{C\to A}
 \right]
 S^{(\text{no D})\dagger}_{C\to A}.
\]
This gives precisely the description obtained by never including $D$ in the
first place, namely $\rho^{(A)}_{CB}$, at least at the level of all probabilities
for the observables in $\widetilde{\cM}_A$. It need not coincide with the
ordinary reduced state $\tr_D\rho^{(A)}_{CBD}$.

This is the sense in which Zanardi's operational approach to subsystems is
relevant. The accessible observables select the effective subsystem structure;
the degrees of freedom to be discarded are those invisible to that algebra
\cite{Zanardi2001,ZanardiLidarLloyd2004}.

\section{Conclusion}

The essence of the apparent third-particle paradox is that two different questions have been conflated:
\begin{enumerate}
\item Does inclusion of the third particle affect the probabilities of the measurements originally available in the $C$-description, after these measurements are transformed to the $A$-description?
\item Does the reduced density operator obtained by tracing out $D$ after the QRF transformation coincide with the density operator obtained by transforming a description in which $D$ was never included?
\end{enumerate}
The answer to the first question is no by construction. The answer to the second question is also no in general. However, our discussion reveals that we should not expect a positive answer: $D$ relative to $C$ and $D$ relative to $A$ are simply different subsystems. The paradox arises only if a negative answer to the second question is mistaken for a positive operational influence in the first question. 

Our analysis shows that there is no third-particle paradox and that subsystems admit a meaningful description relative to QRFs. Different QRF approaches nonetheless treat the issue differently, according to the features each takes as foundational. In the perspectival approach the subsystem structure depends on how many subsystems one considers. This dependence is reflected, for instance, in the mathematical representation of $\widetilde{\mathcal M}_A$, which differs according to whether one includes $D$ alone or the further subsystems $D'$, $D''\dots$ as well. On the perspectival approach, which assumes no external structure, this dependence is unavoidable and natural -- a feature, not a bug. The approach of Ref.~\cite{Krumm2021}, inspired by Dirac quantization, instead drops the identification $O_{AB}^{(C)} \cong \widetilde M_A$ in favor of an embedding independent of the QRF perspective; this is justified if one identifies QRF transformations with a change of representation in a redundant description. The extra-particle approach, which emphasizes the compatibility of the QRF perspective with potential external subsystems, favors an ``incoherent-twirling'' prescription to preserve the subsystem structure upon adding a third system.

Independently of the approach, the general moral is that in QRF physics states, observables and subsystem decompositions must be transformed and interpreted together. A ``system $D$'' may be discarded relative to a given operational algebra, while becoming relevant for a larger algebra of measurements. Thus the right question is not whether ``the third particle'' exists in the description, but what is the relevant, operationally defined subsystem whose probabilities one aims to predict.\\

\noindent \textbf{Acknowledgments.} M.K. thanks M. P. M\"{u}ller, B. Sahdo, P. A. H\"{o}hn, M. Mekonnen, and D. Trillo for discussions. This research was funded in part by the Austrian Science Fund (FWF) [SFB BeyondC F7102, DOI: 10.55776/F71; WIT9503323, DOI: 10.55776/WIT9503323]. For open access purposes, the author has applied a CC BY public copyright license to any author accepted manuscript version arising from this submission. This work was also supported by the European Union (ERC Advanced Grant, QuantAI, No. 101055129). The views and opinions expressed in this article are however those of the author(s) only and do not necessarily reflect those of the European Union or the European Research Council - neither the European Union nor the granting authority can be held responsible for them.

\end{document}